\documentclass[aps,
twocolumn,prX]{revtex4}

\usepackage{amsmath,amsfonts,bm,graphicx}
\begin{document}

\newcommand\<{\langle}
\renewcommand\>{\rangle}
\renewcommand\d{\partial}
\newcommand\LambdaQCD{\Lambda_{\textrm{QCD}}}
\newcommand\tr{\mathrm{Tr}\,}
\newcommand\+{\dagger}
\newcommand\g{g_5}

\title{Tests of Universality in AdS/QCD}
\author{Joshua Erlich}
\affiliation{Department of Physics, College of William and Mary,
Williamsburg, Virginia 23187-8795}
\author{Christopher Westenberger}
\affiliation{Department of Physics, The College of New Jersey, Ewing, New Jersey 08618-1104}


\newcommand\sect[1]{\emph{#1}---}

\begin{abstract}
Estimates of the light hadron masses, decay constants and couplings
in AdS/QCD models are generally 
more accurate than should have been expected.  
Certain predictions based on the AdS/CFT correspondence, 
such as the ratio of the equilibrium viscosity to entropy density,
are universal and therefore provide firm experimental tests of these models.  
Other observables, while not completely universal, may be relatively 
insensitive to model details.
We calculate the dependence of a number of low-energy
hadronic observables on
details of the hard-wall AdS/QCD model.  In particular, we vary 
the infrared boundary conditions, the
5D gauge coupling, and the mass of the field responsible for chiral symmetry
breaking, while holding fixed a small number of observables.  
We also find a generalized Gell-Mann--Oakes--Renner relation which helps to
justify the identification of model parameters with the product of physical
quark mass and chiral condensate as per the AdS/CFT correspondence.

\end{abstract}
\pacs{nn.nn.xx}
\maketitle

\sect{Introduction}%
AdS/QCD is a class of extra-dimensional models in which the
light mesons of quantum chromodynamics (QCD) 
are interpreted as Kaluza-Klein modes.
In recent years, AdS/QCD models have proven surprisingly successful at
reproducing low-energy hadronic observables such as masses and decay
constants.    
Most low-energy observables that have been calculated
in AdS/QCD models agree with experiment at the 10-20\% level.

AdS/QCD was originally motivated by the AdS/CFT correspondence
\cite{Maldacena} in the context of models that confine
\cite{Witten-circle,Polchinski-Strassler} with chiral symmetry breaking
\cite{Klebanov-Strassler,Mateos}, 
although QCD with three colors
certainly does not have a weakly-coupled AdS/CFT dual via this
correspondence.  While it is possible to reproduce a
Regge-like spectrum at high energies \cite{soft},
AdS/QCD models generally make poor predictions
when extrapolated to 
high energies 
(for a recent discussion, see Refs.~\cite{Strassler-unparticle,undelivered}).
These models typically fail above 2 GeV, where
the QCD coupling is small and the AdS/CFT correspondence  
suggests that a classical extra-dimensional description
should not provide a good approximation to QCD.  

Top-down AdS/QCD
models arise from string theory, with the D4-D8 system describing a
gauge theory which is confining with non-Abelian chiral symmetry
breaking as in QCD.  So far all models rooted in string theory
contain states not included in the QCD spectrum.  Furthermore, 
the classical limit
of the AdS/CFT correspondence requires the number of colors, $N$, and the
't Hooft coupling, $g^2N$, to be large.  Nonetheless, predictions
of top-down AdS/QCD models, naively extrapolated to three colors, fare
relatively well when compared to experimental data \cite{D4D8}.
Bottom-up models are also
motivated by the AdS/CFT correspondence, but are more
phenomenological and allow more freedom to match QCD data
 \cite{soft,Strassler-Hong-Yoon,BD,EKSS,DP}.  
The AdS/CFT correspondence
has also motivated extra-dimensional models of electroweak symmetry breaking
\cite{electroweak} and condensed matter systems \cite{condmat}.
With a bit of hindsight, we address the question of why AdS/QCD models
have been so successful for modeling static properties of
the light hadrons.  

Chiral symmetry breaking and QCD sum rules are
the underlying principles of this approach, as in Refs.~\cite{EKSS,DP}.  
Both top-down models such as the D4-D8 system \cite{D4D8} and
bottom-up models such as the hard-wall and soft-wall models
\cite{EKSS,DP,soft} 
generally make similar predictions for low-energy observables,
despite the vastly different effective (4+1)-dimensional spacetime geometries.  
Certain predictions
are known to be exactly universal in models based on the classical limit
of the AdS/CFT correspondence, such as the
shear viscosity-to-entropy density ratio in thermal
equilibrium
$\eta/s=1/(4\pi)$ \cite{eta-s} (which turns out to be in agreement up to a 
factor of a few with experimental estimates for the quark-gluon plasma
from RHIC data).  
Model-independent predictions serve as crucial tests of the AdS/QCD approach.
Conversely, it may be that the success of AdS/QCD models below 2 GeV
is the result of universality of predictions in that regime.

Focusing on hard-wall models, 
we will demonstrate that a number of low-energy observables 
continue to provide an unreasonably good fit to data as the squared mass 
$m_X^2$ of the field
responsible for chiral symmetry breaking in the model is varied over a
reasonable range, 
hinting at universality of AdS/QCD predictions for those observables.  
However, we find some sensitivity to the 5D gauge coupling $g_5$
and the boundary conditions arising from localized kinetic terms for the
5D gauge fields.  Our comparison to experimental data 
suggests that in the space of these 
parameters, there is a saddle point in the root-mean-squared error near
the values determined by
naive application of the AdS/CFT correspondence while matching to the 
ultraviolet (UV).  
The fact that we can successfully extrapolate certain results
from low energies where QCD is strongly coupled, to high energies where
QCD is weakly coupled, remains mysterious.  We will also generalize a
derivation of the
Gell-Mann--Oakes--Renner relation for the pion mass and decay constant
in terms of AdS/QCD model parameters.

\sect{AdS/QCD} 
In this section we review the hard-wall AdS/QCD model.
We would like to identify the towers of radial excitations of mesons in QCD
with Kaluza-Klein modes of fields in an extra dimension.
The spins, parities, and charges of the 
QCD states correspond to the spins, parities, and charges
of the Kaluza-Klein modes.  The charges for now will refer to 
SU(2) isospin representations.  

To reflect the approximate SU(2)$\times$SU(2) chiral symmetry of the
light quarks, we begin with a SU(2)$\times$SU(2) gauge theory in some
(4+1)-dimensional (5D) spacetime background.  
The Kaluza-Klein decomposition of the gauge fields should include a
massive tower of modes, but no zero mode.  The global chiral
symmetry of the (3+1)-dimensional (4D) theory 
is lifted to a higher-dimensional gauge invariance in 
the spirit of hidden local symmetry \cite{hls}, as in the deconstructed
extra-dimensional model of Son and Stephanov \cite{Son-Stephanov}.
The bulk 5D metric and the boundary conditions
for the bulk fields are chosen to preserve 4D Lorentz invariance in the
effective 4D theory.  The metric takes the form 
\begin{equation}
ds^2=e^{-A(z)}\left(dx_\mu dx^\mu-dz^2\right), \label{eq:metric}\end{equation}
where the Greek spacetime indices run from 0 to 3, and the function $A(z)$
describes the warping of the spacetime.  The geometry is manifestly isometric
under 4D Lorentz transformations acting on $x^\mu$ but not $z$.  

A 5D scalar
field charged under the SU(2)$\times$SU(2) gauge group is introduced in order
to dynamically break the gauge group to its diagonal (isospin)
subgroup via the Higgs mechanism \cite{EKSS,DP}.  
The simplest possibility is a
scalar field $X^{ab}$ transforming in the bifundamental representation of 
SU(2)$\times$SU(2), which is assumed to obtain a background profile 
$X_0^{ab}(z)=\delta^{ab}v(z)$, so as to preserve the diagonal SU(2) isospin
subgroup.
Alternatively, or in addition, the boundary conditions
can be chosen such as to break the chiral symmetry 
\cite{Higgsless,Hirn-Sanz}, but we will not consider this possibility here.  
The 5D action is, \begin{eqnarray}
S_{5D}&=&-\frac{1}{8g_5^2}
\int d^5x \,\sqrt{-g}(L_{MN}^aL_{PQ}^a+R_{MN}^aR_{PQ}^a)\, g^{MP}g^{NQ}
\nonumber \\
&&+\tr(|D_M X|^2 -m_X^2 \,|X|^2), \end{eqnarray}
where the capital Latin spacetime indices run from 0 to 4, the gauge 
index $a$ runs from 1 to 3, and $m_X^2$ is the squared
mass of the scalar field $X$.  $L_{MN}^a$ and $R_{MN}^a$ are the field
strength tensors of the two SU(2) gauge group factors.
We normalize the gauge fields as in Ref.~\cite{EKSS}.  

4D parity corresponds to the transformation $x^i\rightarrow -x^i,\ i=1,2,3$,
together with an exchange $L_M^a\leftrightarrow R_M^a$.
The spectrum contains vector mesons, axial vector mesons, 
pseudoscalars, and scalars.  If we allow gravity to fluctuate, then there
are also
spin-two modes that can be identified with a tower of spin-two 
glueballs \footnote{However, in top-down models the glueballs propagate in
the full (9+1)-dimensional spacetime, while the mesons are confined to flavor
branes.}.
With a given geometry (specified by $A(z)$), 
the free parameters in the model are $g_5$, $z_m$, $m_X$,
and the background profile $v(z)$, which should solve the equation of motion
for the field $X$ in the background spacetime.  Since the equation of motion
for $X$ is a second order differential equation, there are two parameters in 
its
solution.  The field $X$ carries the quantum numbers of
the quark bilinear $\overline{q}_L^iq_R^j$.  
The AdS/CFT correspondence identifies the non-normalizable solution
({\em i.e.} the solution with infinite action) with the source for an
operator charged under the chiral symmetry like the field $X$ 
\cite{Witten,GKP}, {\em i.e.}
the quark mass $m_q$ up to an overall normalization; the
normalizable solution
({\em i.e.} the solution with finite action) is related to the expectation
value of the corresponding operator \cite{normalizable1,normalizable2},
in this case the chiral condensate $\sigma=\langle\overline{q}_Lq_R+
\overline{q}_Rq_L\rangle=2{\rm Re}\left(
\langle\overline{q}_Lq_R\rangle\right)$, where 
$q$ here is either the up or down quark.
These identifications follow from the AdS/CFT correspondence, but 
we need not {\em a priori} make these
identifications in order to define the AdS/QCD model.  We also do not
include the gravitational backreaction of the background scalar field
\cite{braneless,Maru}.

There are a number of ways one could choose the function $A(z)$ in the metric.
We could choose the geometry so as to obtain the physical spectrum of
vector mesons, or to at least obtain the linear confining spectrum as in the
soft-wall model \cite{soft}.  Alternatively, although we do not expect
the model to accurately describe QCD at high energies,
we may choose the geometry so
as to best match the high-energy behavior of
certain correlation functions of QCD.
The choice of the AdS geometry in the UV
region (where the warp factor $\exp[-A(z)]=R^2/z^2$ is largest)
is equivalent to an assumption of conformal behavior of QCD at
high energies compared to the mass of the rho meson.  In the hard-wall
model the geometry is assumed for simplicity to be a slice of AdS
between a UV cutoff length $z=\epsilon\rightarrow 0$ and an IR length 
scale $z=z_m$. There is still freedom in
specifying boundary conditions for 5D fields in the IR region of the geometry.
The slice of Anti-de Sitter space with boundary conditions of the form 
$L_{\mu z}(x,z_m)=R_{\mu z}(x,z_m)=0$ for the gauge fields
leads to a spectrum and decay constants which agree 
with a Pad\'e approximation to the two-point function of isospin currents
about a point in the deep Euclidean regime \cite{Shifman}.  
In the spirit of deconstructed extra dimensions, we can
observe the emergence of the radial direction of Anti-de Sitter space as
the order of the Pad\'e approximation becomes large 
\cite{EKL}.
In order to reproduce power corrections to
correlation functions in the UV, including the dependence on the running
coupling, we could add higher dimension operators to the action and modify
the geometry away from Anti-de Sitter space in the UV region 
 \cite{power-corrections}.  However, for simplicity we will focus on 
the simple
hard-wall model in our investigation, with the geometry described above:
a slice of
Anti-de Sitter space described by the metric in Eq.~(\ref{eq:metric}) with
$A(z)=\log(z^2/R^2)$, between $z=\epsilon$ and $z=z_m$.  The AdS curvature
$R$ can be absorbed into the definitions of $g_5$ and $m_X$, so 
we will generally rescale those parameters and set $R=1$.  

If we naively follow
the AdS/CFT correspondence we can fix the squared 
mass $m_X^2$ by its relation to the
conformal dimension of the quark bilinear in the ultraviolet as in 
Refs.~\cite{EKSS,DP}.
According to the AdS/CFT correspondence, the mass of the field $X$  is related
to the scaling dimension $\Delta$
of the corresponding operator, $\overline{q}q$, via
\begin{equation}
m_X^2=\Delta(\Delta-4).\label{eq:mX}\end{equation}  
In the ultraviolet $\Delta=3$, and we would infer $m_X^2=-3$.  However,
the AdS/CFT correspondence does not directly apply, and this identification
of the $X$ mass ignores renormalization effects which tend to reduce the
scaling dimension at lower energies.  We will refer to $\Delta$
as the effective scaling dimension, and we will allow $\Delta$ to vary
in our analysis.  
The solution to the equations of motion for the background $X$-field
takes the form $X(z)=v(z){\mathbf{1}}$, with 
\begin{equation}
v(z)=m_q z^{4-\Delta}+\frac{\sigma}{4(\Delta-2)} z^\Delta. 
\label{eq:v}\end{equation}  
We asume in this analysis that $\Delta>2$.
To the extent
that the model parameters are related to the physical quark
mass and chiral condensate, the factor of $1/(4(\Delta-2))$ multiplying the
condensate $\sigma$ follows from the AdS/CFT correspondence, 
and we assume here that the parameters $m_q$ and
$\sigma$ are real
\footnote{A 
factor of $1/(2\Delta-4)$ appears in the AdS/CFT 
relation between the coefficient
of the normalizable solution $z^\Delta$ and the expectation
value of the corresponding operator. 
(We thank Tom Cohen and Aleksey Cherman for pointing out that
the correct normalization appears in Ref.~\cite{normalizable2}. This
normalization differs
from that in Ref.~\cite{normalizable1} by a subtle rescaling first noticed
by requiring
consistency of the AdS/CFT correspondence with Ward identities \cite{Dan}.)
The additional factor of $1/2$ in Eq.~(\ref{eq:v}) appears
because the chiral condensate is $\langle\overline{q}q\rangle
=\langle \overline{q}_Lq_R+\overline{q}_Rq_L\rangle$, 
while the operator that transforms under the chiral symmetry like 
the field $X$ is
$\overline{q}_Lq_R$ without the addition of its hermitian conjugate.
If we assume that $m_q$ is normalized to be
the quark mass and $\sigma$ has vanishing
imaginary part, then according to the AdS/CFT correspondence
$\sigma$ is the properly normalized chiral condensate.}.

\sect{Correlation Functions and Observables} 
We calculate seven observables: $m_\pi$, $f_\pi$, $m_\rho$, $F_\rho$,
$m_{a_1}$, $F_{a_1}$ and $g_{\rho\pi\pi}$, where $m$'s are meson masses, 
$F$'s are the correpsonding decay constants,
and $g_{\rho\pi\pi}$ describes the coupling of the $\rho$ to pions.
For this analysis we choose to fix the parameters $z_m$, $m_q$ and
$\sigma$ by matching to $m_\rho=775.8$ MeV, $f_\pi=92.4$ MeV, and 
$m_\pi=139.6$ MeV.

Kaluza-Klein modes are the nonvanishing solutions to the classical equations 
of motion subject to the appropriate boundary conditions.  
We work in the axial gauge $L_z^a=R_z^a=0$ and consider transverse fluctuations
with $\partial_\mu L_\perp^\mu=\partial_\mu R_\perp^\mu=0$.
The transverse part of the
diagonal SU(2) gauge fields $V_{\perp\,\mu}^a=(L_{\mu}^a+
R_{\mu}^a)_\perp/2$ satisfy the linearized equation of motion, 
\begin{equation}
\left[\partial_z\left(\frac{1}{z}\partial_z V_\mu^a(x,z)\right)-\frac{1}{z}
\partial_\nu\partial^\nu V_\mu^a(x,z)\right]_\perp=0. \end{equation}
The Kaluza-Klein modes, Fourier transformed in 3+1 dimensions, 
are solutions of the form, \begin{equation}
V_\mu^a(q_n,z)=\varepsilon_\mu^a(q_n)\,\psi_n(z), \end{equation}
where $q_n^2=m_n^2$ is the squared mass of the $n^{\rm th}$ Kaluza-Klein mode, and
$\psi_n(z)$ satisfies the boundary conditions $\psi_n(\epsilon)=0$.  
The wavefunction $\psi_n(z)$ is normalized so that
the kinetic term for the Kaluza-Klein  mode is canonically normalized in the
effective 4D theory.  There is some arbitrariness in the choice of
boundary conditions at $z=z_m$.  In order that the boundary conditions not break the
chiral symmetry, we typically choose gauge-invariant conditions of the form 
$L_{\mu z}^a(x,z_m)=R_{\mu z}^a(x,z_m)=0$
as in Refs.~\cite{EKSS,DP}.  However, this choice is somewhat ad hoc, so in the following section
we will consider a more general class of gauge-invariant boundary conditions,
obtained by the addition of localized gauge kinetic terms.

To calculate decay constants for the vector Kaluza-Klein modes, we couple to 
a background zero-mode for the diagonal isospin gauge field, 
which is normalizable if $\epsilon\neq 0$ \footnote{More precisely, the
zero mode is in the spectrum if Neumann boundary conditions are imposed at
$z=\epsilon$, but the Neumann and Dirichlet conditions become degenerate
while the zero mode decouples from the spectrum as $\epsilon\rightarrow0$.  
Also note that the zero mode is consistent with
our generalized boundary conditions at $z=z_m$.}.
The effective 4D action is obtained by expanding the 5D action in terms of the Kaluza-Klein modes
and integrating over the radial coordinate $z$.  The rho mesons are identified with the lightest
Kaluza-Klein mode of $V_\mu^a$.  The coefficient of the term bilinear in the
Kaluza-Klein mode and the zero mode corresponds to the vacuum-to-one-particle
matrix element of the isospin current, and
defines the decay constant for that mode \cite{D4D8}.  It follows that,
\begin{equation}
F_\rho^2=\frac{1}{g_5^2}\left(\psi_1'(\epsilon)/\epsilon\right)^2.
\end{equation}
The correlator of isospin currents can be written in terms of the masses and
decay constants.
In the approximation of narrow QCD resonances, the correlation function of the product of two
vector currents takes the form, \begin{equation}
i\int d^4x\langle J_\mu^a(x)J_\nu^b(0)\rangle e^{iq\cdot x}=\delta^{ab}
\sum_n \frac{F_n^2}{q^2-m_n^2}\, \left(
g_{\mu\nu}-\frac{q_\mu q_\nu}{m_n^2}\right). 
\label{eq:JJ}\end{equation}
The
bulk-to-boundary propagator is the solution to the Fourier-transformed
classical equation of motion of the form
$V_\mu^a(q,z)=V_\mu^a(q)\,V(q,z)$, subject to
the boundary condition $V(q,\epsilon)=1$.  In terms of $V(q,z)$, the 
sum over modes in Eq.~(\ref{eq:JJ}) can be written 
\cite{Strassler-Hong-Yoon,EKSS,DP}, 
\begin{equation}
\sum_n \frac{F_n^2}{q^2-m_n^2}=\frac{1}{g_5^2}\left.
\frac{\partial_z V(q,z)}{z}\right|_{z=\epsilon}.
 \end{equation}
For large $-q^2$ the correlator is logarithmic as in QCD, and 
matching to the perturbative
result would determine $g_5=2\pi$ \cite{EKSS}.  However, since we
don't expect the model to be valid in the ultraviolet regime, it is reasonable to allow
$g_5$ to vary.  

Decay constants for the axial-vector mesons are determined similarly.  
The axial-vector field is the combination $A_\mu^a=(L_\mu^a-R_\mu^a)/2$.
The transverse part of the axial vector field $A_{\perp\,\mu}^a$ 
satisfies the equation of motion, Fourier transformed in 3+1 dimensions,
\begin{equation}
\left[\partial_z\left(\frac{1}{z}\partial_zA_\mu^a\right)+\frac{q^2}{z}
A_\mu^a-\frac{4g_5^2v(z)^2}{z^3}\,A_\mu^a\right]_\perp=0.
\label{eq:Aperp}\end{equation}
The $a_1$ meson corresponds to the solution satisfying $A_{\perp\,\mu}^a
(\epsilon)=0$
and whatever we choose for the IR boundary conditions at $z=z_m$. 
In the chiral limit, the 
axial current-current correlator has a singularity at $q^2=0$ from the
exchange of pions.  The 
bulk-to-boundary propagator for the axial vector, $A(q,z)$,
satisfies the equation of motion for $A_\perp$ but with UV
boundary condition $A(q,\epsilon)=1$.  The pion decay constant in the
chiral limit takes the form \cite{EKSS,DP}, \begin{equation}
f_\pi^2=-\frac{1}{g_5^2}\left.\frac{\partial_zA(0,z)}{z}\right|_{z=\epsilon}.
\label{eq:fpi}\end{equation}

The longitudinal part of the axial vector field, $A_{\parallel\,\mu}^a
=\partial_\mu
\varphi^a$, mixes with the Goldstone modes $\pi^a(x,z)$
in $X^{ab}=v(z)\delta^{ab}\,\exp(2i\,\pi^aT^a)$, where $T^a$ generate
the broken part of the chiral symmetry group.  In the $A_z=0$ gauge
the equations of motion, Fourier transformed in 3+1 dimensions, are,
\begin{equation}
\partial_z\left(\frac{1}{z}\partial_z\varphi^a\right)+\frac{4g_5^2 v(z)^2}{
z^3}\left(\pi^a-\varphi^a\right)=0, \label{eq:pi1}\end{equation}
\begin{equation}
-q^2\,\partial_z\varphi^a+\frac{4g_5^2 v(z)^2}{z^2}\,\partial_z\pi^a=0.
\label{eq:pi2}\end{equation}
The pion corresponds to the solution satisfying the boundary conditions
$\varphi(\epsilon)=\pi(\epsilon)=0$ and $\varphi'(z_m)=0$. Because
the longitudinal component of the gauge fields do not appear in the 
(3+1)-dimensional components of the field
strengths $L_{\mu\nu}$ and $R_{\mu\nu}$ at the quadratic level, 
the boundary conditions for the longitudinal components
will not be modified when we generalize the boundary conditions in a 
gauge-invariant manner. 
Couplings between mesons are determined from the terms in the
effective 4D action which couple the corresponding Kaluza-Klein modes.  We 
will include
the coupling of the rho to two pions, $g_{\rho\pi\pi}$, in our fits.
In terms of the fields $\pi$ and $\varphi$ the coupling is given by
\cite{EKSS,DP},
\begin{equation}
g_{\rho\pi\pi}=g_5\int dz\,\psi_\rho(z)\left(\frac{\varphi'(z)^2}{g_5^2 z}
+\frac{4v(z)^2(\pi-\varphi)^2}{z^3}\right), \end{equation}
where $\pi$ and $\varphi$ are normalized so that the pion has a canonical
kinetic term in the effective 4D theory.

\sect{Varying Boundary Conditions}
In order to determine how predictions of the seven aforementioned observables
vary as a function of boundary conditions in the IR, we first consider the 
$\rho$ wave functions
subject to $\psi_{\rho}(\epsilon)=0$ and
$a\, m^2_{\rho}\,\psi_{\rho}(z_m)-b\,\d_z\psi_{\rho}(z_m)=0$, where $a,b\in\Re^{+}$. 
The latter boundary condition
is consistent with the general form, \begin{equation}
a \,\partial^\mu V_{\mu\nu}^a(x,z_m)- b\, V_{\nu z}^a(x,z_m)=0, 
\label{eq:bc}\end{equation} 
where $V_{MN}^a$ is the field strength tensor for the diagonal gauge field 
$V_M^a$.  (Our analysis here is linearized in the fields, so there is no
ambiguity from the quadratic parts of the field strength.)  

The boundary conditions (\ref{eq:bc}) arise from the addition of
localized gauge kinetic terms on the IR boundary:
\begin{equation}
S_{5D}\rightarrow S_{5D}-\frac{1}{8g_5^2 z_m}\int d^4x\,\frac{a}{b}
\left(L_{\mu\nu}^cL^{c\,\mu\nu}+R_{\mu\nu}^cR^{c\,\mu\nu}\right).
\end{equation}
In addition to modifying the boundary conditions as in Eq.(\ref{eq:bc}), the
localized kinetic term has the consequence of rescaling the normalization
of the kinetic terms for the Kaluza-Klein modes in 
the effective 4D theory, so to restore
its canonical normalization the $\rho$ wavefunction satisfies,
\begin{equation}
\int_\epsilon^{z_m}\frac{dz}{z}\,\psi_\rho(z)^2+\frac{a}{b\,z_m}\,\psi_\rho(z_m)^2
=1, \end{equation}
and similarly for the $a_1$ wavefunction.

\begin{figure*}
\begin{center}
\includegraphics[scale=1]{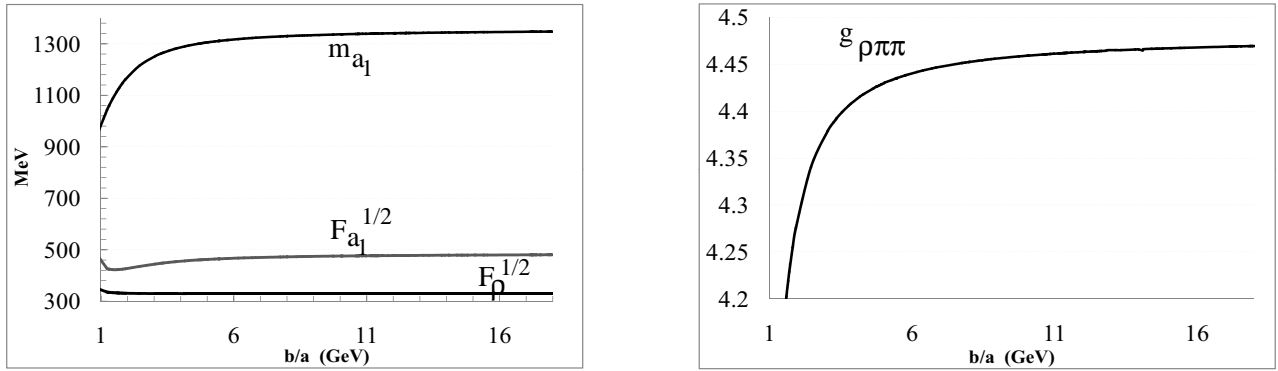}
\end{center}
\caption{Hard-wall model predictions as function of boundary conditions. }\label{fig:bc1}
\end{figure*}

For the time being we fix $g_5=2\pi$ and effective scaling dimension
$\Delta=3$.
For any given value of
$\frac{b}{a}$, we fix $m_{\rho}=775.8$ MeV, and use the IR boundary condition to fix $z_m$.
By fixing $f_\pi=92.4$ MeV, and $m_\pi=139.6$ MeV, we determine
the parameters $m_q$ and $\sigma$. We then calculate
$F_\rho$, $m_{a_1}$, $F_{a_1}$ and $g_{\rho\pi\pi}$ for the chosen value of
$\frac{b}{a}$, which are plotted in Fig.~\ref{fig:bc1}. 
The Neumann boundary condition 
$L_{\mu z}^a=R_{\mu z}^a=0$ corresponds to the limit $b/a\rightarrow\infty$.

\begin{figure}
\begin{center}
\includegraphics[scale=.8]{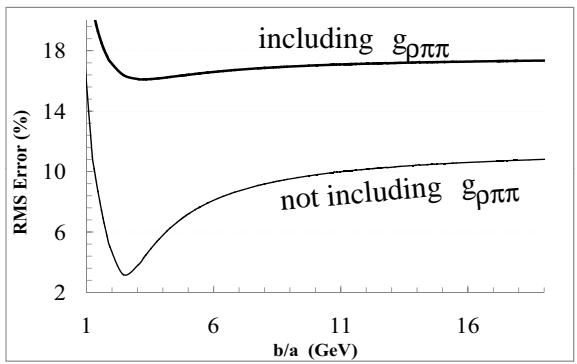}
\end{center}
\caption{RMS Error 
as function of boundary conditions.}\label{fig:bc3}
\end{figure}

We define the RMS error as \cite{EKSS}, \begin{equation}
\varepsilon_{RMS}=\sqrt{\sum_{{\rm observables}}\left(\frac{{\rm theory}-{\rm expt}}{
{\rm expt}}\right)^2\,\frac{1}{n_{{\rm DOF}}}}, \end{equation}
where $n_{{\rm DOF}}$=\# observables $-$ \# parameters.
For the experimental values we took the central values listed in
Ref.~\cite{EKSS}.
In our calculations we
considered $\frac{b}{a}\in\left[0.56,20.0\ {\rm GeV}\right]$, and observed 
that the RMS error of the model (calculated using $F_\rho^{1/2}$, $m_{a_1}$, 
$F_{a_1}^{1/2}$ but not $g_{\rho\pi \pi}$) ranged from values as low as 
$3.2$ percent when $b/a\approx 2.5$ GeV to $30$ percent for smaller values of
$b/a$.  The RMS error is higher if we include
$g_{\rho\pi\pi}$, as
seen in Fig.~\ref{fig:bc3}.  This is
consistent with the intuition that $g_{\rho\pi\pi}$ is more sensistive to
the details of the model, and will also depend on 
higher-dimension operators that have not been included in the action.

\begin{figure*}
\begin{center}
\includegraphics[scale=1,]{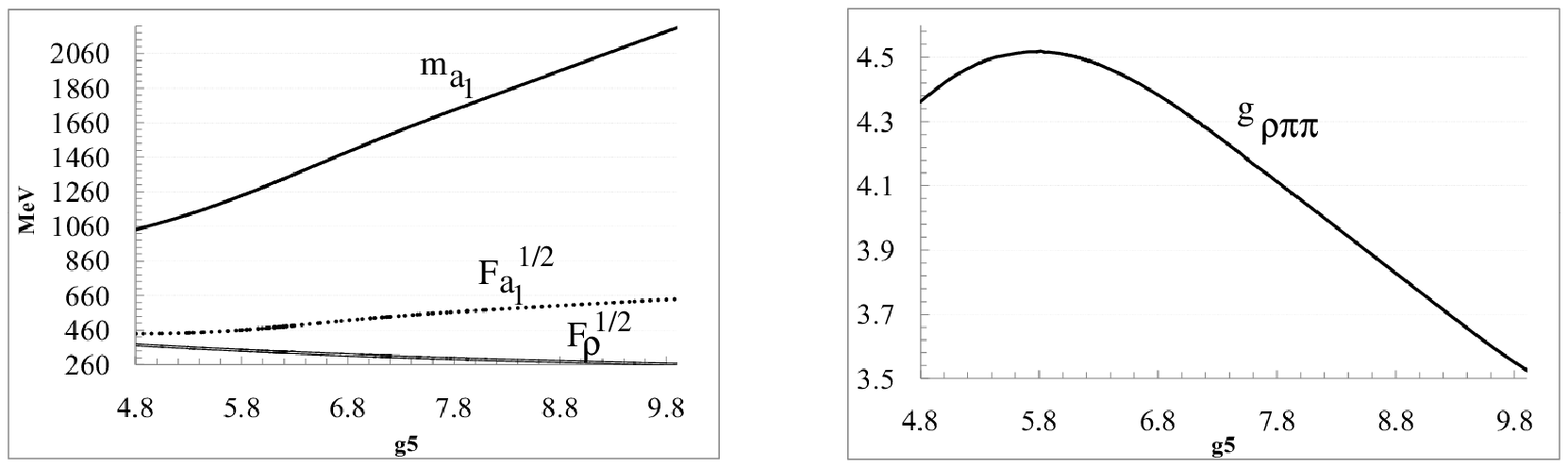}
\end{center}
\caption{Hard-wall model predictions as function of $g_5$. }\label{fig:g51}
\end{figure*}

\begin{figure}
\begin{center}
\includegraphics[scale=.7
]{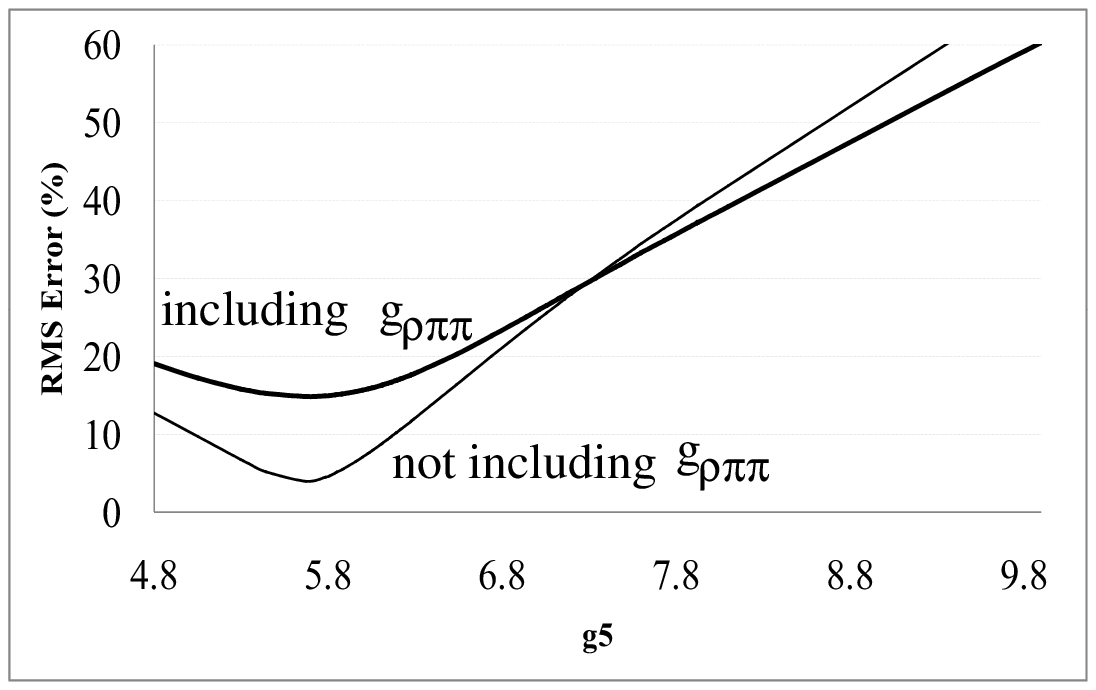}
\end{center}
\caption{RMS Error as function of $g_5$.}\label{fig:g53}
\end{figure}

\sect{Varying $g_5$}
By fixing Neumann boundary conditions in the IR, we now determine how predictions
of the observables vary as a function of the 5D gauge coupling $g_5$. 
Demanding that $m_{\rho}=775.8$ MeV fixes $z_m=1/(323 \text{ MeV})$.
Next, we determine $m_q$ and $\sigma$ by fixing
$f_\pi=92.4$ MeV, $m_\pi=139.6$ MeV.
The observables
$F_\rho$, $m_{a_1}$, $F_{a_1}$ and $g_{\rho\pi\pi}$ are plotted in 
Fig.~\ref{fig:g51}. 
In our calculations we
considered $g_5\in\left[4.8,9.9\right]$, and observed that the RMS error of 
model ranged from values as low as $14.85$ percent to $60.7$ percent and 
is visualized in Fig.~\ref{fig:g53}.
Curiously, the  value of $g_5$ obtained by matching to the ultraviolet,
$g_5=2\pi$, is near a minimum of the RMS error.  This may be used as 
evidence in favor of the naive matching.

\begin{figure*}
\begin{center}
\includegraphics[scale=1]{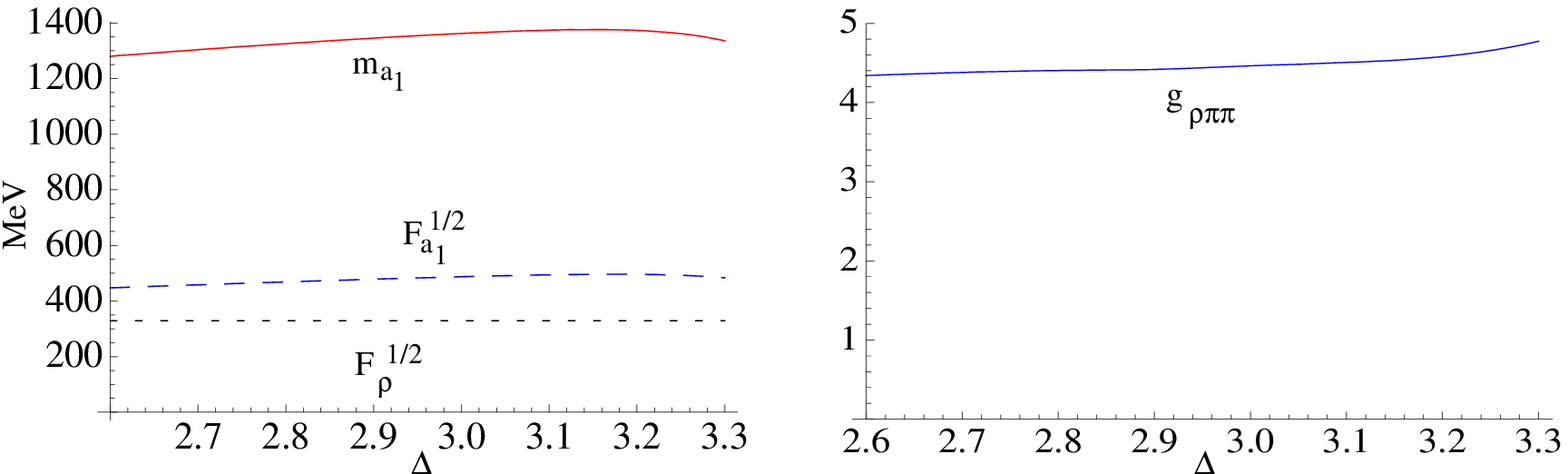}
\end{center}
\caption{Hard-wall model predictions as function of $\Delta$.}\label{fig:Delta1}
\end{figure*}

\begin{figure}
\begin{center}
\includegraphics[scale=0.7,viewport=0 0 500 250,clip]{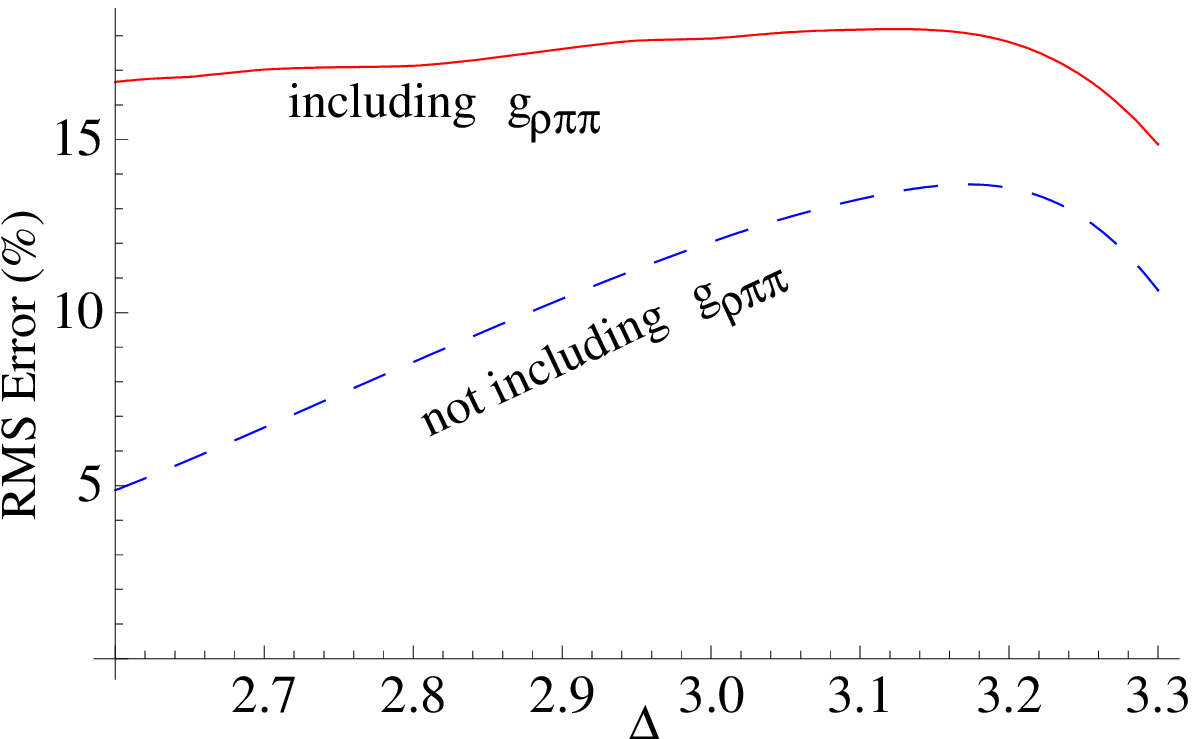}
\end{center}
\caption{RMS Error as function of $\Delta$.  The error is dominated by 
$g_{\rho\pi\pi}$.}
\label{fig:Delta3}
\end{figure}

\sect{Varying $X$ mass}
The mass of the field $X$ is related to the effective scaling dimension
$\Delta$ by Eq.~(\ref{eq:mX}).
Allowing $\Delta$ to vary while fixing
the AdS/QCD predictions for $m_\pi$ and $f_\pi$ introduces a dependence of
the model parameters $m_q$ and $\sigma$ on $\Delta$.  The parameter $z_m$
is determined by the rho mass, which is not affected by $\Delta$.  
Fig.~\ref{fig:Delta1}  
shows the dependence of the
observables on $\Delta$. Fig.~\ref{fig:Delta3} shows how the RMS error
varies with $\Delta$.

\begin{figure}
\begin{center}
\includegraphics[scale=0.7,viewport=0 0 500 250,clip]{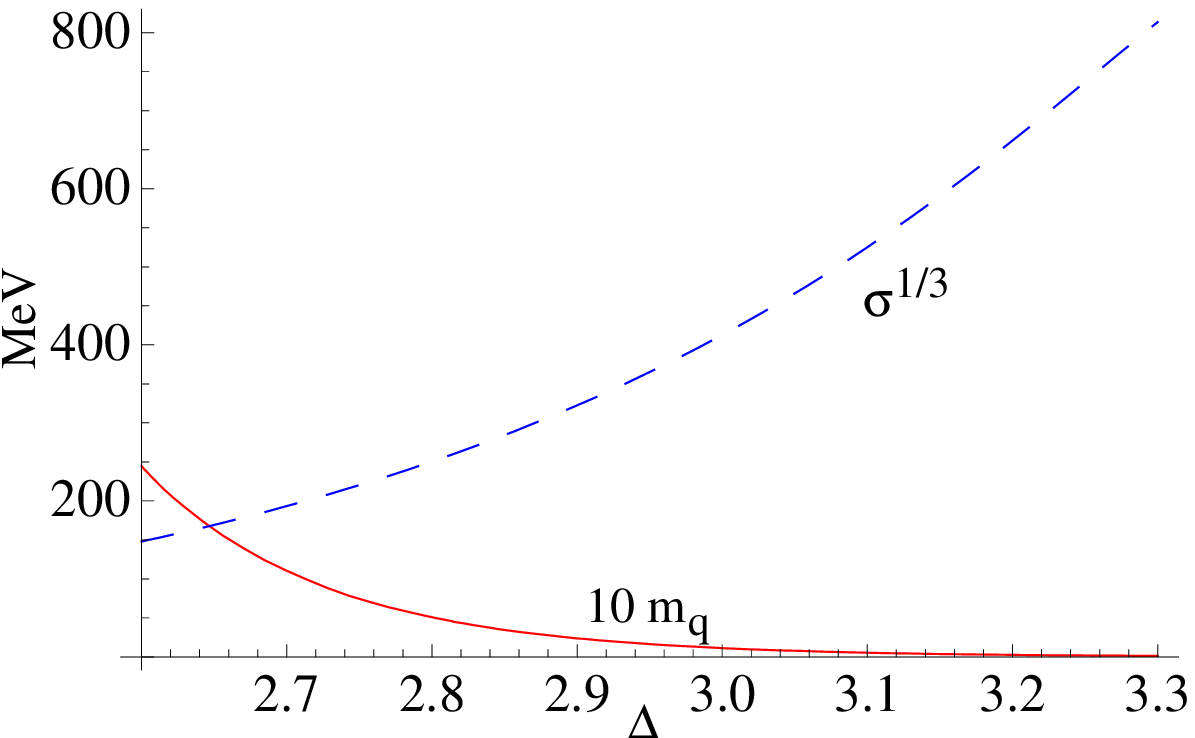}
\end{center}
\caption{The model parameters $10m_q$ and $\sigma^{1/3}$ as a function of
the effective scaling dimension $\Delta$.}\label{fig:GOR1}
\end{figure}

\sect{A Comment on Chiral Symmetry}
In  
Fig.~\ref{fig:GOR1} we plot $10 m_q$ and $\sigma$, fit by $m_\pi$ and
$f_\pi$, as functions of $\Delta$.  (We 
rescale $m_q$ by a factor of ten for visibility in the plot.)  Note that
as the effective scaling dimension 
$\Delta$ decreases from its classical value of three, the parameter
$m_q$ quickly increases.  The
chiral symmetry breaking scale in the model is set by $\sigma$ and $z_m$.
The chiral limit is the regime in which the quark mass is much smaller than the
chiral symmetry breaking scale.  To the extent that $m_q$ represents the light
quark masses, this would seem to imply a deviation from the chiral limit
as $\Delta$ decreases.
It is curious that the fit to data 
disfavors the region of parameter
space $m_X^2\approx-3$, as demonstrated in Fig.~\ref{fig:Delta3}, although
even $m_X^2=-3$ provides a good fit.

\begin{figure}
\begin{center}
\includegraphics[scale=0.7,viewport=0 0 500 250,clip]{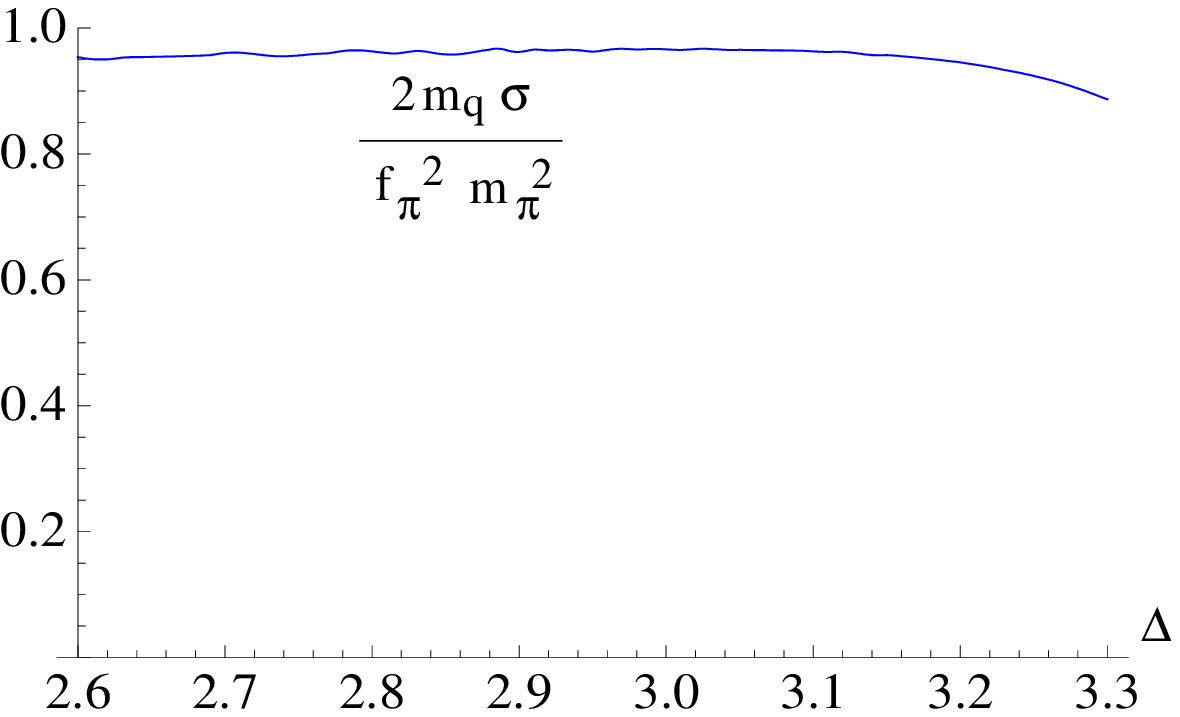}
\end{center}
\caption{Test of the Gell-Mann--Oakes--Renner (GOR) relation. The curve is the
model prediction, with $m_q$ and $\sigma$ fit to the experimental values
of $m_\pi$ and $f_\pi$. The curve is consistently below, but close to,
the approximate value 1 dictated by the GOR relation.}\label{fig:GOR2}
\end{figure}

The Gell-Mann--Oakes--Renner (GOR) relation \cite{GOR}
reflects the pattern of chiral
symmetry breaking.  In terms of the quark mass $m_q$ and the magnitude of
the chiral condensate
$\sigma$, the GOR relation is,
\begin{equation}
m_\pi^2\,f_\pi^2=2m_q\,\sigma. \label{eq:GOR}\end{equation}
In Ref.~\cite{EKSS}, a derivation of the GOR relation was given in which
$m_q$ and $\sigma$ were the parameters of the hard-wall model, with 
effective $\overline{q}q$
scaling dimension $\Delta=3$.  The derivation, which we
generalize below, is independent of the 5D gauge coupling $g_5$ and 
the IR boundary conditions.  However, the normalization of the
GOR relation depends on the $\Delta$-dependent scaling of $\sigma$ as in
Eq.~(\ref{eq:v}).
To see this, we recall the
derivation of the GOR relation from Ref.~\cite{EKSS}.

The transverse part of the axial vector field, $A_{\perp\,\mu}^a$, 
satisfies the equation of motion Eq.~(\ref{eq:Aperp}).
If $q^2=0$ then $\pi(z)={\rm constant}$ 
is a solution to Eq.~(\ref{eq:pi2}), while 
Eq.~(\ref{eq:Aperp}) and Eq.~(\ref{eq:pi1}) become identical.
Hence, to leading order in $m_\pi^2$ we can write the solution for the pion
in terms of the bulk-to-boundary propagator $A(q,z)$, as $\varphi(z)\approx
A(0,z)-1$.  Integrating Eq.~(\ref{eq:pi2}) then gives, \begin{equation}
\pi(z)=m_\pi^2\int_\epsilon^z du\,\frac{u^3}{4v(u)^2}\frac{1}{g_5^2\,u}
\,\partial_uA(0,u).\label{eq:piz}\end{equation}
With $v(z)=(m_q\, z^{4-\Delta}+\sigma/(4\Delta-8)\, z^\Delta)$, the function 
$u^3/v(u)^2$
is negligible except when $u\sim\left(m_q/\sigma\right)^{\frac{1}{2\Delta-4}}$. 
For such small $u$ we can replace
$u\rightarrow \epsilon$ in the remainder of the integrand, which we
recognize as $-f_\pi^2$ from Eq.~(\ref{eq:fpi}).  For $z\gg 
\left(m_q/\sigma\right)^{\frac{1}{2\Delta-4}}$, 
the integral becomes constant, with $\pi(z)\rightarrow-1$ and $\varphi(z)=
A(0,z)-1$ solving the equations of motion.
With these
approximations, the integral in Eq.~(\ref{eq:piz})
becomes, \begin{equation}
\pi(z_m)\approx -\frac{m_\pi^2\,f_\pi^2}{2m_q\,\sigma}\approx-1,
\end{equation}
which finally yields the GOR relation, Eq.~(\ref{eq:GOR}).
We test the GOR relation in Fig.~\ref{fig:GOR2}.
In addition to the dependence of the model parameters
on the effective scaling dimension $\Delta$
in the analysis above, it was recently stressed in 
Ref.~\cite{Cohen} that the operator dual to the field $X$ via the AdS/CFT
correspondence is rescaled from 
$\overline{q}_Lq_R$ by an $N_c$-dependent
factor.  However, that factor would rescale $m_q$ and $\sigma$ oppositely
such that the product $m_q\,\sigma$ is invariant, and the GOR relation
as expressed above continues to hold.

The classical value of $\Delta$ using the 
AdS/CFT correspondence, $\Delta=3$, is near a local maximum of the RMS error.
The error is dominated by $g_{\rho\pi\pi}$, which improves if we
allow $\Delta$ to increase significantly
from its classical value, contrary to the intuition
from asymptotic freedom.  On the other hand, as mentioned earlier
the coupling $g_{\rho\pi\pi}$ 
is sensitive to higher-dimension operators in the 5D action which have not
been included in this analysis,
whereas the quadratic observables ({\em i.e.} 
masses and decay constants) are not.

\sect{Conclusions}
We have studied the dependence of low-energy hadronic observables on 
parameters of the hard-wall AdS/QCD model.  The 5D gauge coupling
and $X$-field mass obtained by matching to the ultraviolet, $g_5=2\pi$ and
$m_X^2=-3$, seem to lie near a saddle point of the RMS error of the model.
The value $g_5=2\pi$ is preferred by data, but the model is in even
better agreement with data if $m_X^2$ is away from its UV-matched value.
We have reproduced
the Gell-Mann--Oakes--Renner (GOR) relation in terms of the model parameters
$m_q$ and $\sigma$, defined with an $m_X$-dependent scaling
consistent with the AdS/CFT correspondence.  
We found some dependence of the observables on the
choice of boundary conditions for the 5D gauge fields, and the model can 
be made to better fit experimental data by an appropriate choice of boundary
conditions.
Our results help to justify the matching of AdS/QCD to  current correlators 
in the ultraviolet,
and support some aspects of AdS/QCD universality.

\sect{Acknowledgments}
We thank Aleksey Cherman, Tom Cohen, Leandro Da Rold,
Hovhannes Grigoryan, Ami Katz, and Matthew Reece for useful discussions. 
The work of JE was supported in part by NSF grants PHY-0504442 and
PHY-0757481.  Part of this work was completed during the Research
Experience for Undergraduates program at the College of William and Mary.
Some of this work was reported at the 8th conference on Quark Confinement
and the Hadron Spectrum \cite{Confinement8}.

\end{document}